\documentclass[aps,prl,twocolumn,showpacs,superscriptaddress,groupedaddress]{revtex4}  
\usepackage{amssymb}   
\usepackage{amsmath}
\usepackage{graphicx}

\newcommand{\ord}{\begin{cal}O\end{cal}}

\def\beq{\begin{equation}}
\def\eeq{\end{equation}}
\def\bsp#1\esp{\begin{split}#1\end{split}}

\newcommand{\asb}[0]{\frac{\alpha_s}{\pi}}
\newcommand{\pasb}[1]{\left(\frac{\alpha_s}{\pi}\right)^{#1}}
\newcommand{\Dplus}[1]{\left[\frac{\log^{#1}(1-z)}{1-z}\right]_+}
\newcommand{\DplusOne}[0]{\left[\frac{\log(1-z)}{1-z}\right]_+}
\newcommand{\DplusZero}[0]{\left[\frac{1}{1-z}\right]_+}

\begin{document}

\begin{flushleft} 
\mbox{IPPP/14/17, DCPT/14/34, ZU-TH 07/14, CERN-PH-TH/2014-042, FERMILAB-PUB-14-052-T}
\end{flushleft}

\title{Higgs boson gluon-fusion production at threshold in N$^3$LO QCD}

\author{Charalampos Anastasiou}
\affiliation{Institute for Theoretical Physics, ETH Z\"urich,
  8093 Z\"urich, Switzerland}

\author{Claude Duhr}
\affiliation{Institute for Particle Physics Phenomenology,
University of Durham, Durham, DH1 3LE, U.K.} 

\author{Falko Dulat}
\affiliation{Institute for Theoretical Physics, ETH Z\"urich,
  8093 Z\"urich, Switzerland}

\author{Elisabetta Furlan}
\affiliation{
Fermilab, Batavia, IL 60510, USA
} 

\author{Thomas Gehrmann}
\affiliation{Physik-Institut, Universit\"at Z\"urich, 
Winterthurerstrasse 190, 8057 Z\"urich, Switzerland}

\author{Franz Herzog}
\affiliation{
CERN Theory Division, CH-1211, Geneva 23, Switzerland
}

\author{Bernhard Mistlberger}
\affiliation{Institute for Theoretical Physics, ETH Z\"urich,
  8093 Z\"urich, Switzerland}

\date{\today}

\begin{abstract}
We present the cross-section for the threshold production of the Higgs boson at hadron-colliders  
at next-to-next-to-next-to-leading order (N$^3$LO) in perturbative QCD. We present an analytic expression for the 
partonic cross-section at threshold and the impact of these corrections on the numerical estimates 
for the hadronic cross-section at the LHC. With this result we achieve a major milestone 
towards a complete evaluation of the cross-section at N$^3$LO which will reduce the theoretical 
uncertainty in the determination of the strengths of the Higgs boson interactions. 
\end{abstract}

\maketitle

High precision theoretical predictions for the production rate of the Higgs boson are  
crucial in the study of the recently discovered particle from the ATLAS and 
CMS collaborations~\cite{HiggsDiscovery} and for inferring the existence of phenomena beyond the 
Standard Model. With the collection of further data at the upgraded LHC, 
the theoretical uncertainty for the gluon-fusion cross-section will become 
soon dominant. It is thus highly timely to improve the theoretical accuracy  
of the cross-section predictions. 

The quest for accurate Higgs boson cross-sections has been long-standing 
and it is paralleled with major advances in perturbative QCD. 
State-of-the-art calculations of the gluon-fusion cross-section (for a review, see Ref.~\cite{ihixs} 
and references therein) comprise 
next-to-leading-order (NLO) QCD corrections in the full Standard-Model theory, 
next-to-next-to-leading order (NNLO) QCD corrections as an expansion in 
inverse powers of the top-quark mass $1/m_t$, two-loop electroweak corrections and mixed 
QCD/electroweak corrections. To improve upon the present accuracy, the most significant 
correction is expected from the N$^3$LO QCD contribution
 in the leading order of the $1/m_t$ expansion.

Universal factorization of radiative corrections due to soft emissions, 
as well as knowledge of the three-loop splitting functions~\cite{Moch:2004pa}, have made possible the 
derivation of logarithmic contributions to the cross-section beyond 
NNLO~\cite{Moch:2005ky}.  However, further progress in determining  the N$^3$LO correction 
can only be achieved by direct evaluation of the Feynman diagrams at 
this order. 

Recently, there was rapid progress in this direction.
The required three-loop matrix-elements have been computed in Ref.~\cite{formfactor}.  
The partonic cross-sections for the production of a Higgs boson in association with three partons 
was computed in Ref.~\cite{triplereal}, while the two-loop matrix-elements for the production of a Higgs boson 
in association with a single parton and the corresponding two-loop soft current were computed in Ref.~\cite{Gehrmann:2011aa} and Ref.~\cite{Duhr:2013msa}. Corrections due to one-loop amplitudes for a Higgs boson in association with a single parton 
were evaluated in Refs.~\cite{Anastasiou:2013mca}, and counter-terms due to ultraviolet~\cite{Wilson,UV} 
and initial-state collinear divergencies were computed in Refs.~\cite{NNLOXsec}.  The N$^3$LO Wilson coefficient and the renormalization constants of the operator in 
the effective theory where the top quark is integrated out 
have been computed in Refs.~\cite{Wilson}. 
Although all these contributions are separately divergent in four dimensions, a finite cross-section 
can be obtained by combining them with the remaining one-loop matrix elements for the production of the Higgs boson in 
association with two partons. 

The purpose of this Letter is to complete the computation of all matrix-elements integrated over loop momenta and phase-space 
which are required at N$^3$LO in the limit of Higgs production at threshold. We present the fully analytic result for the first term in the threshold expansion of the gluon-fusion cross-section at N$^3$LO,
 and we use this result to estimate the impact of N$^3$LO corrections to the inclusive Higgs production cross-section at threshold. Our result is the first calculation of a hadron collider observable 
 at this order in perturbative QCD. 

The Higgs production cross-section takes the form 
\begin{equation}
\label{eq:sigma}
\sigma = \sum_{ij} \int dx_1\, dx_2\, f_i(x_1)\,f_j(x_2)\, \hat{\sigma}_{ij}(m_H^2,x_1\,x_2\,s)\,,
\end{equation} 
where  $\hat{\sigma}_{ij}$ are the partonic cross-sections for producing a Higgs boson from partons $i$ and $j$, $f_i(x_1)$ and $f_j(x_2)$ are the corresponding parton distribution functions, and $m_H^2$ and $s$ denote the mass of the Higgs boson and the hadronic centre-of-mass energy, respectively. We work in an effective theory where the top quark has been integrated out, and the Higgs boson couples directly to the gluons via the effective operator
\begin{equation}
\mathcal{L}_{\textrm{eff}} = -\frac{1}{4v}\,C(\mu^2)\,H\,G_{\mu\nu}^a\,G_a^{\mu\nu}\,,
\end{equation}
where $v\simeq 246$ GeV is the vacuum expectation value of the Higgs field and $C(\mu^2)$ is the Wilson coefficient, given as a perturbative expansion in the $\overline{\text{MS}}$-renormalized strong coupling constant $\alpha_s \equiv \alpha_s(\mu^2)$ evaluated at the scale $\mu^2$. Up to three loops, we have~\cite{Wilson}
\begin{align}
&C(\mu^2) = -\frac{\alpha_s}{3\,\pi}\,\Bigg\{1+\frac{11}{4} \asb\\
&\,+\pasb{2} \Bigg[\frac{19 }{16}\,L_t+\frac{2777}{288}+N_F \left(\frac{1}{3}\,L_t-\frac{67}{96}\right)\Bigg]\nonumber\\
&\,+\pasb{3} \Bigg[\frac{897943 }{9216}\,\zeta _3+\frac{209 }{64}\,L_t^2+\frac{1733}{288}\,L_t-\frac{2892659}{41472}\nonumber\\
&\,\qquad+N_F \left(-\frac{110779 }{13824}\,\zeta _3+\frac{23 }{32}\,L_t^2+\frac{55 }{54}\,L_t+\frac{40291}{20736}\right)\nonumber\\
&\,\qquad+N_F^2 \left(-\frac{1}{18}\,L_t^2+\frac{77 }{1728}\,L_t-\frac{6865}{31104}\right)\Bigg]+\ord(\alpha_s^4)\Bigg\}\,,\nonumber
\end{align}
with $L_t = \log(\mu^2/m_t^2)$ and $N_F$ the number of active light flavours. 

The partonic cross-section itself admits the perturbative expansion
\beq
\label{eq:sigma_partonic}
\hat{\sigma}_{ij}(m_H^2,\hat{s}) = \frac{\pi\,C(\mu^2)^2}{v^2\,V^2}\,
\sum_{k=0}^\infty\pasb{k}\,\eta_{ij}^{(k)}(z)
\,,
\eeq
with $z\equiv m^2_H/\hat{s}$ and $V=N^2-1$, where $N$ denotes the number of colours. The coefficients $\eta_{ij}^{(k)}(z)$ are known explicitly through NNLO in perturbative QCD~\cite{nnlo}.

If all the partons emitted in the final state are soft, we can approximate the partonic cross-sections by their threshold expansion,
\beq
\eta_{ij}^{(k)}(z) = \delta_{ig}\,\delta_{jg}\,\hat{\eta}^{(k)}(z) + \ord(1-z)^0\,.
\eeq
Note that the first term in the threshold expansion, the so-called \emph{soft-virtual} term, only receives contributions from the gluon-gluon initial state. Soft-virtual terms are linear combinations of a $\delta$ function and plus-distributions, 
\beq
\int_0^1dz\left[\frac{g(z)}{1-z}\right]_+f(z) \equiv \int_0^1dz\,\frac{g(z)}{1-z}\,[f(z)-f(1)]\,.
\eeq
Through NNLO, we have~\cite{nnlo,nnlosoft}
\begin{align}
&\hat{\eta}^{(0)}(z) = \delta(1-z)\,,\\
&\hat{\eta}^{(1)}(z) = 2 \,C_A\,\zeta_2 \, \delta(1-z) + 4\, C_A\, \DplusOne\,,\\
&\hat{\eta}^{(2)}(z) = \delta(1-z)\, \Bigg\{C_A^2\,\left(\frac{67 }{18}\,\zeta _2-\frac{55 }{12}\,\zeta _3-\frac{1}{8}\,\zeta_4+\frac{93}{16}\right) \nonumber\\
&+N_F \left[C_F\,\left(\zeta _3-\frac{67}{48}\right)-C_A\,\left(\frac{5 }{9}\,\zeta _2+\frac{1}{6}\,\zeta _3+\frac{5}{3}\right) \right]\Bigg\}\nonumber\\
&+\DplusZero \Bigg[ C_A^2\,\left(\frac{11}{3}\, \zeta _2+\frac{39 }{2}\,\zeta _3-\frac{101}{27}\right)\\
&\qquad+N_F\,C_A\,\left(\frac{14}{27}-\frac{2}{3}\,\zeta _2\right)\Bigg]\nonumber\\
&+\DplusOne \Bigg[C_A^2\,\left(\frac{67}{9}-10\, \zeta _2\right) -\frac{10 }{9}\,C_A\, N_F\Bigg]\nonumber\\
&+\Dplus{2} \left(\frac{2}{3}\,C_A\, N_F-\frac{11}{3}\,C_A^2\right)\nonumber\\
&+ \Dplus{3}\,8\, C_A^2\nonumber\,.
\end{align}
In this expression $\zeta_n$ denotes the Riemann zeta function, $C_A=N$ and $C_F=V/(2N)$. 
For simplicity renormalization and factorisation scales are set equal to the Higgs mass, $\mu_R=\mu_F=m_H$. 

The main result of this Letter is the next term in the perturbative expansion, N$^3$LO, of the cross-section for the threshold production of a Higgs boson. All ingredients necessary to compute $\hat{\eta}^{(3)}(z)$ have recently become available. Each of these contributions is 
individually divergent. Adding up all the contributions, and including the counter-terms necessary to remove the ultraviolet and infrared divergences, all the poles in the dimensional regulator $\epsilon$ cancel, 
leaving a finite remainder in the Laurent expansion, which, for $\mu_R=\mu_F=m_H$, is given by, 
\allowdisplaybreaks
\begin{widetext}
\begin{align}\label{eq:SVN3LO}
&\hat{\eta}^{(3)}(z) = \delta(1-z)\,\Bigg\{
C_A^3\,\left(-\frac{2003 }{48}\,\zeta _6+\frac{413}{6}\, \zeta _3^2-\frac{7579 }{144}\,\zeta _5+\frac{979}{24}\, \zeta _2\, \zeta _3-\frac{15257 }{864}\,\zeta _4-\frac{819 }{16}\,\zeta _3+\frac{16151 }{1296}\,\zeta _2+\frac{215131}{5184}\right) \\
&\quad+N_F \Bigg[C_A^2\,\left(\frac{869 }{72}\,\zeta _5-\frac{125}{12}\, \zeta _3\, \zeta _2+\frac{2629 }{432}\,\zeta_4+\frac{1231 }{216}\,\zeta _3-\frac{70 }{81}\,\zeta _2-\frac{98059}{5184}\right) \nonumber
\\
&\quad\qquad+C_A\, C_F\,\left(\frac{5 }{2}\,\zeta _5+3 \zeta _3 \zeta _2+\frac{11}{72}\,\zeta _4+\frac{13 }{2}\,\zeta _3-\frac{71}{36}\,\zeta _2-\frac{63991}{5184}\right) 
+C_F^2\,\left(-5 \zeta _5+\frac{37 }{12}\,\zeta _3+\frac{19}{18}\right) \Bigg]\nonumber\\
&\quad+N_F^2\, \Bigg[C_A\,\left(-\frac{19 }{36}\,\zeta _4+\frac{43 }{108}\,\zeta _3-\frac{133 }{324}\,\zeta _2+\frac{2515}{1728}\right) + C_F\,\left(-\frac{1}{36}\,\zeta _4-\frac{7}{6}\,\zeta _3-\frac{23}{72}\, \zeta _2+\frac{4481}{2592}\right)\Bigg]
\Bigg\}\nonumber\\
&+\DplusZero\,\Bigg\{
C_A^3\,\left(186\, \zeta _5-\frac{725}{6}\, \zeta _3\, \zeta _2+\frac{253 }{24}\,\zeta _4+\frac{8941}{108}\, \zeta _3+\frac{8563 }{324}\,\zeta _2-\frac{297029}{23328}\right)+ N_F^2\,C_A\,\left(\frac{5 }{27}\,\zeta _3+\frac{10 }{27}\,\zeta _2-\frac{58}{729}\right) \nonumber\\
&\quad+N_F \Bigg[C_A^2\,\left(-\frac{17 }{12}\,\zeta _4-\frac{475 }{36}\,\zeta _3-\frac{2173}{324}\, \zeta _2+\frac{31313}{11664}\right) +C_A\, C_F\,\left(-\frac{1}{2}\,\zeta _4-\frac{19 }{18}\,\zeta _3-\frac{1}{2}\,\zeta _2+\frac{1711}{864}\right) \Bigg]
\Bigg\}\nonumber\\
&+\DplusOne\,\Bigg\{
C_A^3\,\left(-77 \zeta _4-\frac{352 }{3}\,\zeta _3-\frac{152 }{3}\,\zeta _2+\frac{30569}{648}\right) +N_F^2\,C_A \,\left(-\frac{4 }{9}\,\zeta _2+\frac{25}{81}\right)\nonumber \\
&\quad+N_F\, \Bigg[C_A^2\,\left(\frac{46 }{3}\,\zeta _3+\frac{94 }{9}\,\zeta _2-\frac{4211}{324}\right) +C_A\, C_F\,\left(6\, \zeta _3-\frac{63}{8}\right) \Bigg]
\Bigg\}\nonumber\\
&+\Dplus{2}\,\Bigg\{C_A^3\,\left(181\, \zeta _3+\frac{187 }{3}\,\zeta _2-\frac{1051}{27}\right) +N_F \Bigg[C_A^2\,\left(-\frac{34 }{3}\,\zeta _2+\frac{457}{54}\right) +\frac{1}{2}\,C_A\, C_F\Bigg]-\frac{10}{27}\,N_F^2\, C_A \Bigg\}\nonumber\\
&+\Dplus{3}\,\Bigg\{C_A^3\,\left(-56\, \zeta _2+\frac{925}{27}\right) -\frac{164}{27}\,N_F\, C_A^2 +\frac{4}{27}\,N_F^2\, C_A \Bigg\}\nonumber\\
&+\Dplus{4}\,\left(\frac{20}{9}\,N_F\, C_A^2 -\frac{110 }{9}\,C_A^3\right) 
+\Dplus{5}\,8\, C_A^3\nonumber\,.
\end{align}
\end{widetext}
Equation~\eqref{eq:SVN3LO} is the main result of this Letter. While the terms proportional to plus-distributions were previously known~\cite{Moch:2005ky}, we complete the computation of $\hat{\eta}^{(3)}(z)$ by the term proportional to $\delta(1-z)$, which includes in particular all the three-loop virtual corrections. 

Before discussing some of the numerical implications of Eq.~\eqref{eq:SVN3LO}, we have to make a comment about the 
validity of the threshold approximation. As we will see shortly, the  plus-distribution terms show a complicated pattern 
of strong cancellations at LHC energies; the formally most singular terms cancel against sums of less singular ones. Therefore, 
exploiting the formal singularity hierarchy of the terms in the partonic cross-section 
does not guarantee a fast-converging expansion for the hadronic cross-section.  
Furthermore, the definition of threshold corrections in the integral of Eq.~\eqref{eq:sigma} is  
ambiguous, because the limit of the partonic cross-section at threshold is not  affected 
if we multiply the integrand by a function $g$ such that $\lim_{z \to 1} g(z) =1$,
\begin{equation}\label{eq:sigma_normalization}
\int dx_1\, dx_2\, \left[ f_i(x_1)\,f_j(x_2) z g(z) \right] \lim_{z\to 1} \left[ \frac{\hat{\sigma}_{ij}(s,z)}{z g(z)} \right] \,.
\end{equation}
It is obvious that Eq.~\eqref{eq:sigma_normalization} has the same formal accuracy in the threshold expansion, provided that $\lim_{z \to 1} g(z) =1$. As we will see in the following, this ambiguity has a substantial numerical implication, and thus presents an obstacle for obtaining precise predictions. We note however that by including in the future further corrections in the threshold expansion, this ambiguity will be reduced.

Bearing this warning in mind, we present some of the numerical implications of our result for $g(z) = 1$.
For $N=3$ and $N_F=5$, the coefficients of the distributions in Eq.~\eqref{eq:SVN3LO} take the numerical values
\begin{align}
\hat{\eta}^{(3)}(z)&\simeq 
\delta(1-z)\,1124.308887\ldots          & (\to   5.1\%) \nonumber\\
& + \DplusZero\,1466.478272\ldots   & (\to  -5.85\%) \nonumber\\
& - \DplusOne\,6062.086738\ldots    & (\to  -22.88\%)\nonumber\\
& + \Dplus{2}\,7116.015302\ldots      & (\to  -52.45\%)\nonumber\\
& -\Dplus{3}\,1824.362531\ldots        & (\to  -39.90\%)\nonumber\\
 &-\Dplus{4}\,230                                 & (\to  20.01\%) \nonumber \\
& +\Dplus{5}\,216\,.                                & (\to  93.72\%) \nonumber
\end{align}
\begin{figure}[!th]
\includegraphics[width=0.45\textwidth]{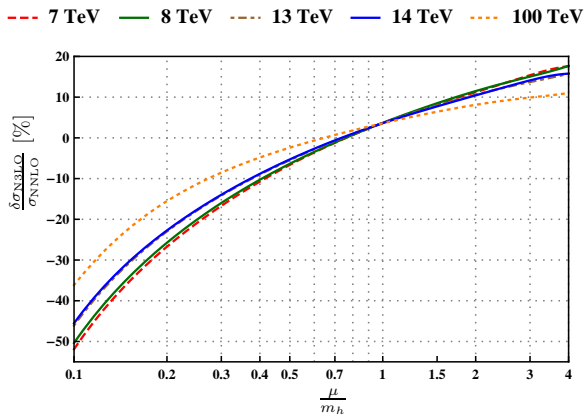}
\caption{
\label{fig:plot}
Percent change from the {\tt ihixs} cross-section at NNLO $\sigma_{\rm NNLO}$ to the N$^3$LO cross-section estimate at threshold for $\sqrt{s} = 7, 8, 13$ and $14$ TeV respectively, as a function of the scale $\mu = \mu_R = \mu_F$.}
\vspace{-0.5cm}
\end{figure}
In parentheses we indicate the correction that each term induces to the hadronic cross-section normalized 
to the leading order cross-section at a center of mass energy of 14 TeV. 
The ratio is evaluated with the MSTW NNLO~\cite{mstw} parton densities and $\alpha_s$ at scales $\mu_R=\mu_F=m_H$ 
in the numerator and denominator. We also factorize the Wilson coefficient at all orders, as in Eq.~\eqref{eq:sigma_partonic},  
in both numerator and denominator, and it cancels in the ratio. We find
that the pure N$^3$LO threshold correction is 
approximately $-2.27\%$ of the leading order. We observe that the $\delta$-term which we computed for the first time in this publication is as large as the sum of 
the plus-distribution terms which were already known in the literature and cancels almost completely against them for 
$\mu_R=\mu_F=m_H$. We note, however, that by choosing a different functional form for the function $g(z)$ in Eq.~\eqref{eq:sigma_normalization}, the conclusion can be substantially different.
For example, by choosing $g(z)=1, z, z^2, 1/z$ 
we find that the threshold correction to the hadronic cross-section at N$^3$LO normalized to the leading order cross-section 
is $-2.27 \%,8.19 \%, 30.16 \%, 7.73 \%$ respectively.

In Fig.~\ref{fig:plot} we present the percentual change of the N$^3$LO threshold corrections to an existing Higgs cross-section 
estimate based on previously known corrections (NNLO, electroweak, quark-mass effects)  in ${\tt ihixs}$~\cite{ihixs} and the settings of Ref.~\cite{ihixs8tev}. The new N$^3$LO correction displayed in this plot includes the full logarithmic dependence on the renormalization and factorization scales, as they can be predicted from renormalization group and DGLAP evolution,  the Wilson coefficient at N$^3$LO and the threshold limit of Eq.~\eqref{eq:SVN3LO}. The function $g(z)$ of Eq.~\eqref{eq:sigma_normalization} is fixed to unity. $\sigma_{\rm NNLO}$ and $\delta \sigma_{\textnormal{N}^3\textnormal{LO}}$ are defined after expanding the product of the Wilson coefficient and the partonic cross-sections in $\alpha_s$.  We conclude that N$^3$LO corrections are important for a high precision estimation of the Higgs cross-section. 

Our result of  the N$^3$LO cross-section at threshold 
demonstrates that it is, in principle,  possible to calculate all loop and phase-space 
integrals required for N$^3$LO QCD corrections for hadron collider processes, albeit in a kinematic limit. 
With this publication, we open up a new era in precision phenomenology  which promises the computation 
of full N$^3$LO corrections for Higgs production and other processes in the future.

{\bf Acknowledgements:}
We are grateful to A. Lazopoulos and S. B\"uhler for their help with {\tt ihixs}.  
Research supported by the Swiss National Science Foundation (SNF) under 
contracts 200021-143781 and  200020-149517, the European Commission through 
the ERC grants ``IterQCD'', ``LHCTheory" (291377) and ``MC@NNLO" (340983) as well as
the FP7 Marie Curie Initial Training Network ``LHCPhenoNet"  (PITN-GA-2010-264564),
and by the U.S. Department of Energy 
under contract no. DE-AC02-07CH11359.


\end{document}